# Molecular dynamics simulations of the lattice thermal conductivity of CuInTe$_2$

J. Wei, H. J. Liu[*], L. Cheng, J. Zhang, P. H. Jiang, J. H. Liang, D. D. Fan, J. Shi

*Key Laboratory of Artificial Micro- and Nano-structures of Ministry of Education and School of Physics and Technology, Wuhan University, Wuhan 430072, China*

The lattice thermal conductivity of thermoelectric material CuInTe$_2$ is predicted using classical molecular dynamics simulations, where a simple but effective Morse-type interatomic potential is constructed by fitting first-principles total energy calculations. In a broad temperature range from 300 to 900 K, our simulated results agree well with those measured experimentally, as well as those obtained from phonon Boltzmann transport equation. By introducing the Cd impurity and Cu vacancy, the thermal conductivity of CuInTe$_2$ can be effectively reduced to further enhance the thermoelectric performance of this chalcopyrite compound.

Ternary I-III-VI chalcopyrite compounds (I=Cu, Ag, III= Al, Ga, In and VI=S, Se, Te) have been identified as good thermoelectric materials [1, 2, 3, 4] due to their lower thermal conductivity. Among them, the thermoelectric performance of CuInTe$_2$ compound [5, 6] has been widely reported. For the *p*-type CuInTe$_2$, the efficiency of thermoelectric conversion, which is characterized by the so-called *ZT*, can reach 1.18 at 850 K [7]. The relatively larger *ZT* value of CuInTe$_2$ is due to the larger Seebeck coefficient ($S$) and moderate electrical conductivity ($\sigma$). In order to further improve the thermoelectric performance of CuInTe$_2$, many efforts have been devoted to optimize the power factor ($S^2\sigma$). Alternatively, one can take some strategies to suppress the lattice thermal conductivity ($\kappa_l$). It is worth noting that the thermal conductivity of CuInTe$_2$ at room temperature is relatively larger than those of state-of-the-art thermoelectric materials such as $Bi_2Te_3$ (~1 W/mK) [8] and filled skutterudites (~3 W/mK) [9]. For example, Liu *et al.* [7] and Kuhn *et al.* [10] have successfully synthesized the $CuInTe_2$ compound and the measured thermal conductivities are 6 W/mk and 3.4 W/mk, respectively. By doping the $CuInTe_2$ compound with Ni or Hg, Kucek *et al.* [11,12] found that the thermal conductivity decreases with increasing doping concentration. Using melting-annealing method, Cheng *et al.* [13] synthesized a series of $CuIn_{1-x}Cd_xTe_2$ compounds and the measured room temperature thermal conductivity is 3.96 W/mK for $x$=0.1. By isoelectronic substitution of element In (Ga) for Ga (In) in the $CuMTe_2$ ($M$ = Ga, In) lattices, Li *et al.* [14] showed that the thermal conductivity of $CuGa_{0.36}In_{0.64}Te_2$ could be reduced to 0.71 W/mK at 701 K. Moreover, by spark plasma sintering, Chen *et al.* [15] fabricated the $CuInTe_2$/graphene composites and found that the thermal conductivity can be successfully decreased. By addition of $TiO_2$ nanofibers into the $CuInTe_2$ system, Luo *et al.* [16] reported that a significant reduction (36%) of the lattice thermal conductivity can be realized and thus greatly enhance the *ZT* value. It should be mentioned that most of these works are experimental and the measured thermal conductivity of $CuInTe_2$ may be quite different in different samples. In this regards, a theoretical investigation is highly desired to not only clarify the existing discrepancy, but also serve a guide to related experimental studies. In this work, the lattice thermal conductivity of $CuInTe_2$ is calculated by using equilibrium molecular dynamic (MD) simulations. We also consider the effect of Cd impurity and Cu vacancy on the thermal conductivity, and the optimized defect concentration is predicted to further enhance the thermoelectric performance of this chalcopyrite compound.

---

[*] Author to whom correspondence should be addressed. Electronic mail: phlhj@whu.edu.cn

The lattice thermal conductivity is predicted by using MD simulation combined with the Green-Kubo theory. To accurately describe the interatomic interactions of the $CuInTe_2$ compound, we choose the Morse-type potential in the form of $D\left[\left(1-\exp\left(-a\left(r-r_0\right)\right)\right)^2-1\right]$, where $D$ is the depth of the potential well, $a$ is the bond elasticity, $r$ represents the interatomic separation, and $r_0$ is the corresponding equilibrium distance. These potential parameters can be determined by fitting the first-principles total energy calculations [17, 18, 19]. During the MD, 200 ps constant temperature simulation (*NVT*) is used to equilibrate the system and then 10 ns constant energy simulation (*NVE*) is carried out to calculate the heat current for every time step of 5 fs. To double-check the reliability of our MD simulation, we have also calculated the thermal conductivity by solving the phonon Boltzmann transport equation [20].

In Table I, we list all the fitted Morse potential parameters for the pristine $CuInTe_2$, which can almost reproduce the total energies obtained from first-principles calculations, as indicated in Figure 1 with respect to the variation of lattice constant. It should be mentioned that the equilibrium lattice constants we found ($a = b = 6.28$ Å and $c = 12.59$ Å) are very close to the experimental values ($a = b = 6.20$ Å and $c = 12.44$ Å) [21]. Using these fitted Morse potential parameters, we have calculated the elastic constants ($C_{ij}$) of $CuInTe_2$ compound. As shown in Table II, our results agree well with those obtained previously from both ionic charge theory [22] and *ab-initio* study [23], which further confirms that our fitted Morse potential is very effective to describe the interatomic interactions in the $CuInTe_2$ compounds.

Based on the accurate interatomic potential, we can now perform MD simulations to predict the lattice thermal conductivity of $CuInTe_2$. The advantage of this method is that it is much faster than other approaches and can thus effectively deal with very large systems or systems with doping. Due to the use of periodic boundary conditions, one need to consider the finite-size effect [24, 25] when using the Green-Kubo theory to calculate the lattice thermal conductivity. We thus perform convergence test by computing the lattice thermal conductivity of $CuInTe_2$ at different lattice sizes (containing 128, 432, 1024, 2000, 3456 and 5488 atoms). In addition, the integration time is also carefully tested. With increasing system, we find that the room temperature thermal conductivity is well converged at a $6 \times 6 \times 6$ supercell containing 3456 atoms. As the phonon mean free path at room temperature is larger than that at high temperature, it is reasonable to use such large supercell to calculate the thermal conductivity at high temperature region.

Figure 3 plots the MD calculated lattice thermal conductivity of $CuInTe_2$ as a function of temperature ranging from 300 K to 900 K, where the inset shows the above-mentioned convergence test of size effect. Here, each value of thermal conductivity is obtained from the average of three independent MD simulations with different Maxwell-Boltzmann velocity distributions. As can be seen from the figure, the lattice thermal conductivity of CuInTe$_2$ decreases with increasing temperature and roughly shows a $T^{-1}$ law, which indicates that the Umklapp process plays a main contribution to the thermal conductivity. Moreover, we find that the MD predicted thermal conductivities are close to those measured experimentally by Li *et al*. [14]. To further verify our MD results, the lattice thermal conductivity of CuInTe$_2$ have also been calculated by solving the phonon Boltzmann transport equation, and we find good agreement between them in the whole temperature region considered.

In order to further reduce the thermal conductivity of CuInTe$_2$, one can take different strategies according to the detailed scattering mechanism such as phonon-phonon scattering, phonon-carriers scattering, grain boundary scattering, and defects scattering. As the impurity and vacancy scattering are relatively easier to be realized experimentally, we will focus on them in the following discussions. In our previous work [26], we have demonstrated that the In atoms can be substituted by atoms with less valence electrons to further optimize the power factor of CuInTe$_2$. It should be noted that such kind of impurity will also induce defect scattering of phonons. As a result, the lattice thermal conductivity of CuInTe$_2$ can be simultaneously reduced. Here, we take Cd doping as an example. Similar to the pristine CuInTe$_2$, we first construct an appropriate Morse potential for the doped system CuIn$_{1-x}$Cd$_x$Te$_2$, where the fitted potential parameters are summarized in Table III. In our MD simulations, the Cd-doped system is generated by randomly replacing In atoms with Cd atom in the supercell. We see from Figure 3 that Cd doping can indeed reduce the lattice thermal conductivity of CuInTe$_2$, which is mainly caused by the atomic mass difference and radius fluctuation between In and Cd atoms. Such effect is more pronounced at room temperature, where we find that the thermal conductivity of CuInTe$_2$ is reduced by 37% when the Cd concentration increases from 0.0 to 0.06. In contrast, there is only a slight reduction of about 4% at 900 K. As high frequency phonons only contribute to the thermal conductivity in a significant way at high temperature region due to stronger phonon-phonon scattering, the relatively weaker influence on the thermal conductivity at high temperature indicates that Cd atoms mainly suppress the low frequency phonons while have less effect on the high frequency phonons.

We next consider the effect of Cu vacancy. Figure 4 plots the lattice thermal conductivity of $Cu_{1-x}InTe_2$ as a function temperature from 300 K to 900 K. We see that the Cu vacancy basically has similar behavior as that of the Cd-doped system. The only difference is that the vacancy is more efficient in suppressing the thermal conductivity than Cd doping. For example, at the same temperature of 900 K, a vacancy concentration of 0.05 leads to a thermal conductivity reduction of 21%, while it is only decreased by 4% for the case of Cd doping, even with a higher concentration of 0.06. It should be mentioned that Kosuga *et al.* [27] reported that with the increasing concentration of Cu vacancy, the electrical conductivity of $CuInTe_2$ increases while the Seebeck coefficient decreases simultaneously. Although larger concentration of Cu vacancy means lower lattice thermal conductivity, the Seebeck coefficient will become smaller. In another word, there is a strong interdependence of the transport coefficients (electrical conductivity, Seebeck coefficient and lattice thermal conductivity), which suggests that we should choose an appropriate concentration of Cu vacancy to find a tradeoff between the electronic and phonon transport so that the thermoelectric performance of $CuInTe_2$ can be enhanced.

Taking into account with the electronic transport properties of $CuInTe_2$ compound [26], we can now investigate the effect of defect (Cd doping or Cu vacancy) concentration on the thermoelectric performance. Figure 5 shows the calculated *ZT* values as a function of defect concentration at 850 K. For the Cd-doped system ($CuIn_{1-x}Cd_xTe_2$), the *ZT* value increases with Cd concentration *x*, and reach a peak of 1.55 at *x*=0.04, followed by a slow decay. Similar picture can be found for the system with Cu vacancy ($Cu_{1-x}InTe_2$). At a slightly lower defect concentration of *x*=0.035, we see the maximum *ZT* value of 1.75 can be achieved, which is obviously larger than that of Cd-doped system. It should be emphasized that in a wide range of defect concentration from $x$=0.01 to $x$=0.07, the *ZT* values of $Cu_{1-x}InTe_2$ system are always larger than those of $CuIn_{1-x}Cd_xTe_2$ system, which is in consistent with experimental results [13, 27]. More importantly, our predicted *ZT* value of 1.75 is obviously larger than the best of those reported experimentally [16], which suggests that there is still room to significantly enhance the thermoelectric performance of $CuInTe_2$ if the defect concentration can be fine controlled.

We thank financial support from the National Natural Science Foundation (Grant No. 11574236 and 51172167) and the "973 Program" of China (Grant No. 2013CB632502).

**Table I** Fitted parameters in the Morse potential for the CuInTe$_2$ compound.

| Bond | D (eV) | a (Å$^{-1}$) | r$_0$ (Å) | Cutoff (Å) |
|---|---|---|---|---|
| Cu-Cu | 0.19 | 1.26 | 4.49 | 5.00 |
| Cu-In | 0.21 | 1.16 | 4.49 | 4.70 |
| Cu-Te | 0.44 | 1.05 | 2.63 | 3.50 |
| In-In | 0.21 | 1.12 | 4.49 | 5.00 |
| In-Te | 0.46 | 1.07 | 2.88 | 3.50 |
| Te-Te | 0.18 | 1.26 | 4.34 | 5.00 |

**Table II** Elastic constants $C_{ij}$ (in unit of GPa) of the CuInTe$_2$ compound, calculated using the fitted Morse potential, and are compared with other theoretical results.

| | $C_{11}$ | $C_{12}$ | $C_{13}$ | $C_{33}$ | $C_{44}$ | $C_{66}$ |
|---|---|---|---|---|---|---|
| ionic charge theory [22] | 80.6 | 50.4 | 58.9 | 93.1 | 34.1 | 31.0 |
| *ab-initio* study [23] | 61.0 | 36.9 | 40.5 | 67.3 | 27.2 | 24.8 |
| this work | 61.3 | 36.7 | 34.6 | 64.1 | 25.9 | 25.7 |

**Table III** Fitted parameters in the Morse potential for the CuIn$_{1-x}$Cd$_x$Te$_2$ compound.

| Bond | D (eV) | a (Å$^{-1}$) | r$_0$ (Å) | cutoff (Å) |
|---|---|---|---|---|
| Cd-Cu | 0.50 | 1.26 | 4.42 | 5.00 |
| Cd-In | 0.21 | 1.16 | 4.42 | 5.00 |
| Cd-Te | 0.41 | 1.05 | 2.85 | 3.50 |

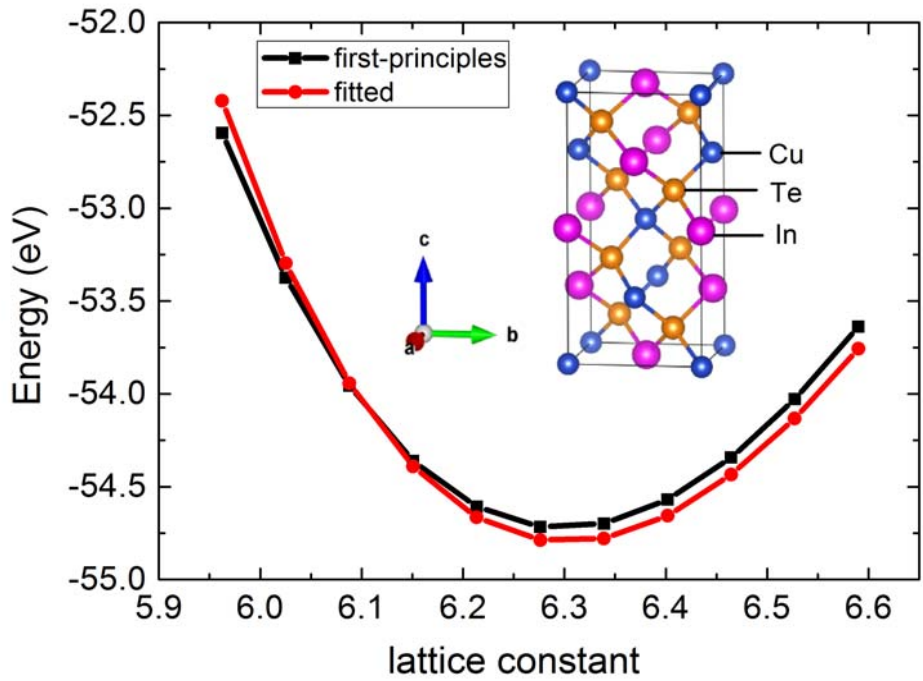

**Figure 1** (Color online) Variation of total energy of CuInTe$_2$ compound with lattice constant, the results from first-principles total energy calculations and those with fitted Morse potential are both shown for comparison. The inset shows the crystal structure of CuInTe$_2$.

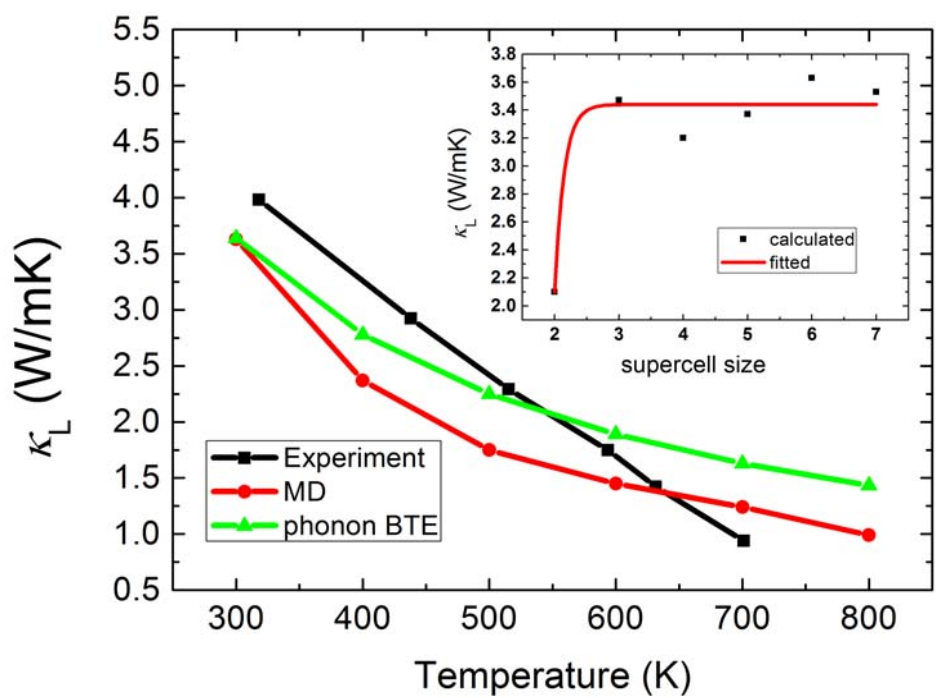

**Figure 2** (Color online) The MD predicted lattice thermal conductivity of CuInTe$_2$ compound as a function of temperature. The experimental results and those predicted by solving the phonon Boltzmann transport equation are both shown for comparison. The inset shows the convergence test of size effect.

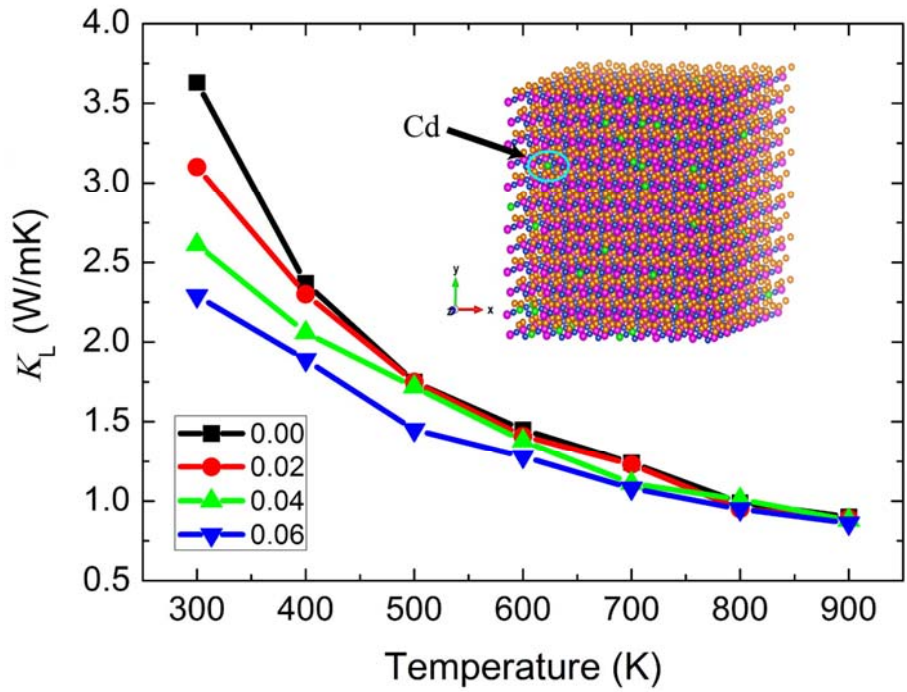

**Figure 3** (Color online) The MD predicted lattice thermal conductivity of CuIn$_{1-x}$Cd$_x$Te$_2$ ($x$=0~0.06) compounds. The inset illustrates the simulation box with the In atom substituted by Cd atom.

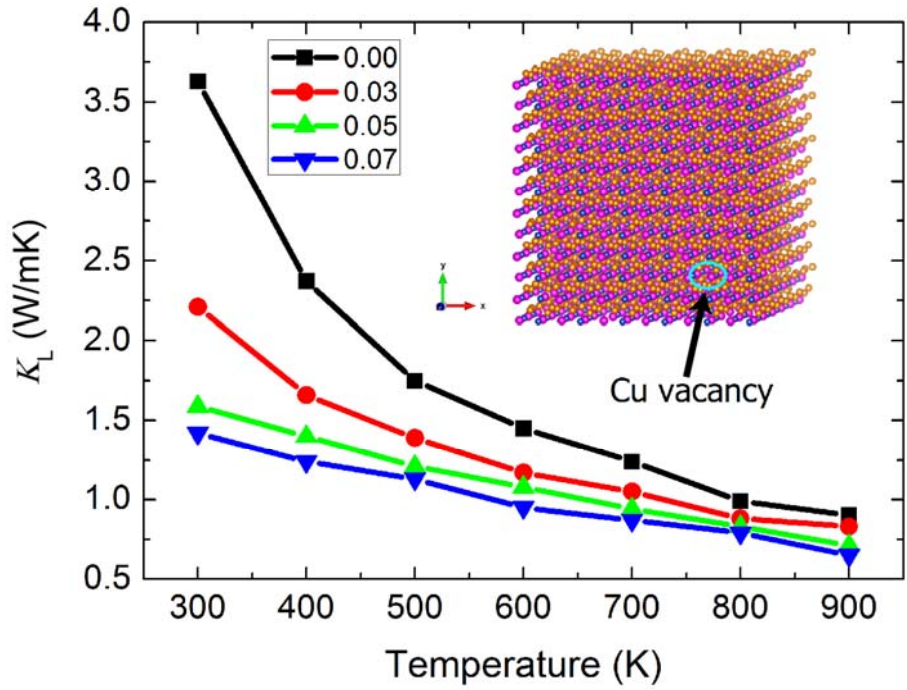

**Figure 4** (Color online) The MD predicted lattice thermal conductivity of $Cu_{1-x}InTe_2$ ($x$=0~0.07) compounds. The inset illustrates the simulation box with Cu vacancy.

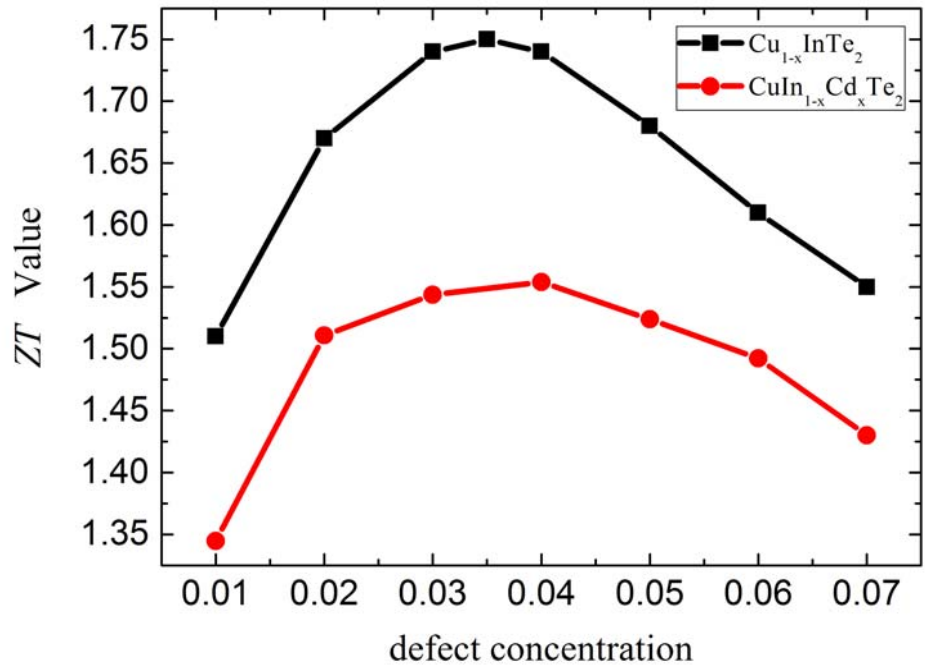

**Figure 5** (Color online) The calculated *ZT* values of CuInTe$_2$ with Cd impurity and Cu vacancy, plotted as a function of defect concentration at 850 K.